# Decomposition of the Inequality of Income Distribution by Income Types—Application for Romania

**Tudorel Andrei [1,\*], Bogdan Oancea [2], Peter Richmond [3], Gurjeet Dhesi [4] and Claudiu Herteliu [1,\*]**

[1] Department of Statistics and Econometrics, Bucharest University of Economic Studies, Bucureşti 010374, Romania

[2] Department of Economic and Administrative Sciences, University of Bucharest, Bucureşti 050107, Romania; bogdan.oancea@faa.unibuc.ro

[3] School of Physics, Trinity College Dublin, Dublin 2, Ireland; peter_richmond@ymail.com

[4] School of Business, London South Bank University, London SE1 0AA, UK; dhesig@lsbu.ac.uk

\* Correspondence: andrei.tudorel@csie.ase.ro or andreitudorel@yahoo.com (T.A.); hertz@csie.ase.ro or claudiu.herteliu@gmail.com (C.H.); Tel.: +40-722-455-586 (C.H.)

**Abstract:** This paper identifies the salient factors that characterize the inequality income distribution for Romania. Data analysis is rigorously carried out using sophisticated techniques borrowed from classical statistics (Theil). Decomposition of the inequalities measured by the Theil index is also performed. This study relies on an exhaustive (11.1 million records for 2014) data-set for total personal gross income of Romanian citizens.

## 1. Introduction

Many analyses of income distributions have been made over the years by economists and econophysicists (for example: [1–14]). Of special interest in all these analyses are income distribution and income inequality (see for example: [15–27]). Several distribution functions have been proposed to describe the income distribution, of which the lognormal-Pareto [28] and exponential-Pareto distributions [29,30] best fit the empirical data.

Romanian income distribution has been little studied. Using wage data from a social security database for a county in Romania, Derzsy et al. [31] showed that in the upper tail, the distribution follows a Pareto law with a coefficient of 2.5, in the range of low and middle incomes, they found that an exponential distribution fits the data. Similar results were obtained by Oancea et al. [29] using tax records data for the entire population that received an income in Romania in 2013. Other studies of the Romanian income distribution [32,33] used survey data and showed that income inequality in Romania has grown over time.

In this paper, we use tax records data for 2014 in Romania to study income inequality using the Theil index [34–36]. While there are other widely used measures of inequality like the Gini index or the Lorenz curve, we are mainly interested in studying the extent to which income inequality can be explained by different subgroups of populations. In this case, the advantages of the decomposable measures of inequality make the Theil index the ideal candidate to be considered in our analysis [37]. We decompose the total income of a person by the source of income in wages and non-wage income (here we include social transfers, unemployment benefits etc.) and grouped the population by the source of income. We then study the decomposition of the Theil index by population subgroups starting with the guidelines presented in [37]. However, while Shorrocks [37] used a decomposition for disjoint groups, our population groups overlap since individuals can earn income from multiple sources during a year. Decomposition of inequality by income sources was studied in several papers using either the classical method, where the Theil index can be seen as a weighted average of inequality within subgroups, or the inequality between those subgroups examined via regression-based methods: [38–41] or [42–45]. In this paper, we introduce a new decomposition of the Theil index where the population groups overlap.

## 2. Problem Presentation

The income of an individual in a population has three possible sources: salary, capital, or other sources like pensions, unemployment benefits, and social assistance. In Romania, every person who was registered as having earned an income during 2014 earned money from one, two, or three of these sources. In these conditions, the total population having an income during 2014 was divided into the following seven categories of persons (Figure 1): persons who had their income only from a single source of income (G1—salaries, G2—capital, and G3—other sources of income), persons who earned income from two income categories (G4—salaries and other sources, G5—salaries and capital income, and G6—capital income and other sources of income), and persons who earned money from all three income sources (G7). Figure 1 shows the nature of the different incomes associated with individuals in the different categories, G1 to G7.

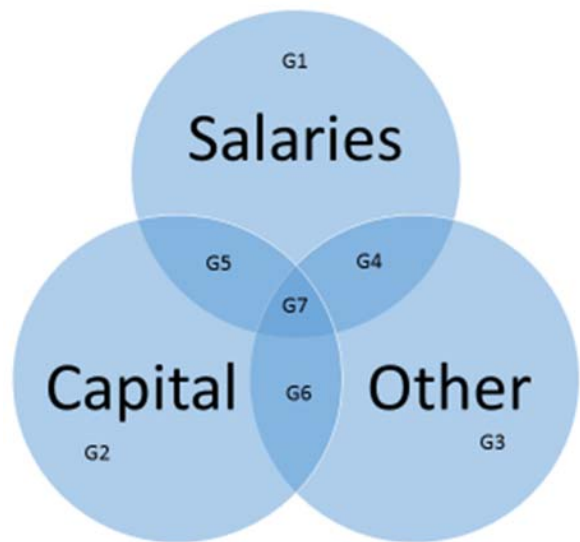

**Figure 1.** The nature of the different incomes associated with individuals in the different categories, G1 to G7.

Thus, we divide the total population into seven disjoint groups:

$$P = \cup_{i=1}^{7} G_i, G_i \cap G_j = \emptyset, \forall i \neq j, i, j = 1, \dots, 7 \tag{1}$$

Under these conditions, we study the following:

- To what extent is the inequality of the distribution of the total income of the population influenced by the distribution of income on the seven classes of persons. In this case, we use a decomposition of the Theil index calculated for the entire population depending on the inequality of distribution of income among the seven groups of people and the differences that exist between the seven groups. In this case, the decomposition of the Theil index corresponds to the case where the groups are disjoint [37]. The total inequality is explained by the factors that act at the level of the groups and factors that differentiate the groups of employees;
- For each group for which there are at least two income sources, the inequality of income distribution is measured by the inequality of income distribution on each source of income. In this case, because the decomposition relationship used at the previous point can no longer be used, we propose another relationship for this decomposition: the total inequality of income distributions can be decomposed into three components: the first component highlights the differences that are at the level of each data series, the second component highlights the differences between the averages of the data series, and the last term highlights the interaction between the factors.

### 3. Data Series

We use gross annual income computed from tax records data at the individual level for 2014. We distinguish between three parts of the total income: wages, capital income, and other sources of income. The currency used for all incomes in the current paper is the Romanian "leu"—RON. The first part of the income can be attributed to labor (domestic and abroad). The second part, capital income, comes from dividends, interest on deposits, rents, real estate transfers, etc. The third part of the income comes from pensions, social assistance, unemployment benefits, income from agricultural labor, freelance activities, and intellectual property rights. Our data sets have 11 million records and were processed using R software [46,47].

The structure of incomes and across the population shows that (i) 44% of people who earned wages earned 56% of the total income of the population; (ii) 23% of people have earned capital income and these represent 19.5% of the total income of the population; (iii) 33% of people also earned other types of income, accounting for 24.4% of the total income of the population; (iv) Figure 2 shows the shares of income and the number of people in the seven groups—these results show that there are significant differences between the two data series; and (v) the significant differences between income shares and the number of people in the seven groups are materialized by different yearly average earnings of persons belonging to one of the seven groups of persons (Figure 3).

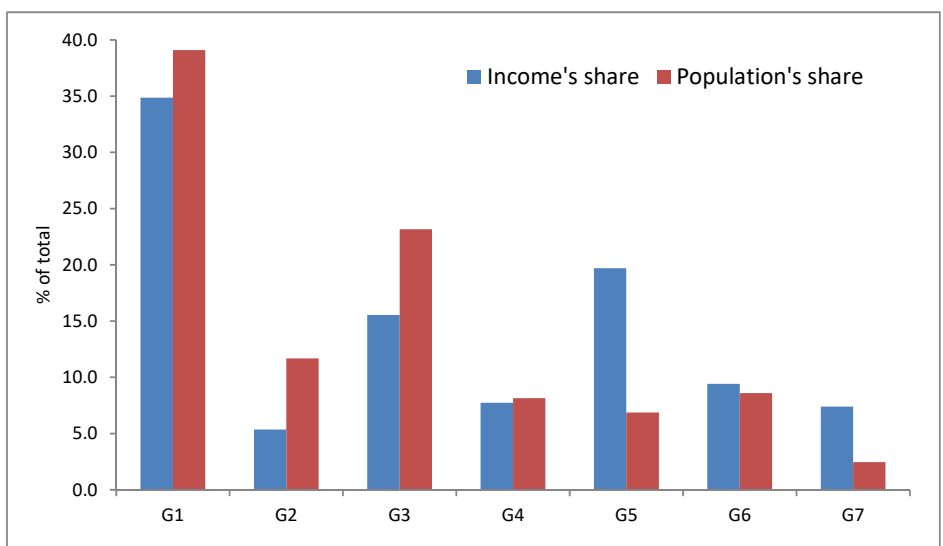

**Figure 2.** The share of the income and of the population who earned an income during 2014.

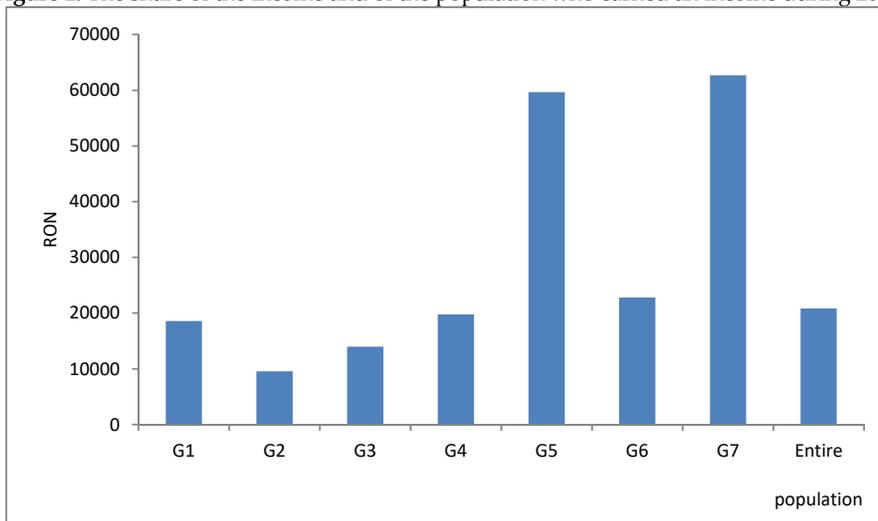

**Figure 3.** Average annual income per person for the entire population and for the seven groups of population.

Table 1 presents the characteristics of the total incomes of the population, both for total and for the three sources of income. The characteristics are evaluated for the whole population and for the seven groups. The results in Table 1 allow the following comments to be made:

(i) At the level of the total population, there are significant differences in the distribution of the income obtained by the source of income (Figure 4);

(ii) There is a different concentration of the income from the three sources. Figure 5 shows the ratio between top 1%/bottom 99% for the total population, the seven groups, and the income from the three sources of income (we define this ratio as the ratio of the sum of incomes in the 99–100% centile to the sum of incomes in the 0–1% centile);

(iii) The total income of the population is thus divided on the three sources of income: 56.0%—wages, 19.5%—capital and 24.4%—other income;

(iv) The distribution of people who have earned income from at least one source of income on the three sources is as follows: 44% obtained at least income from wages, 23%—at least capital income and 33%—at least other income.

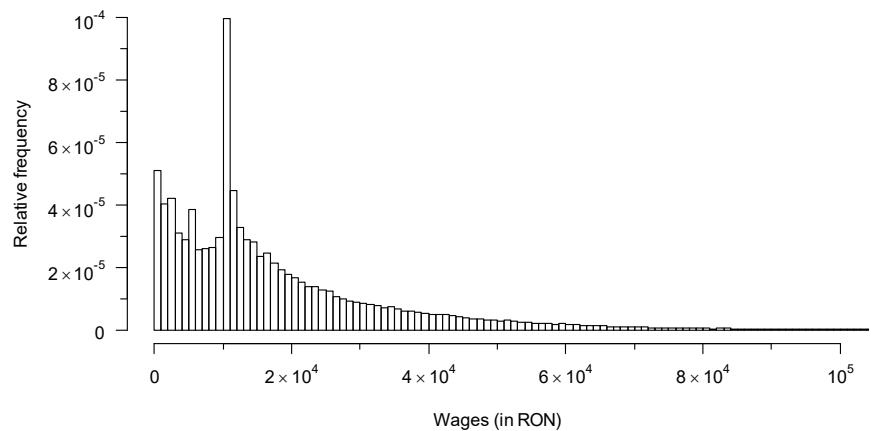

(A)

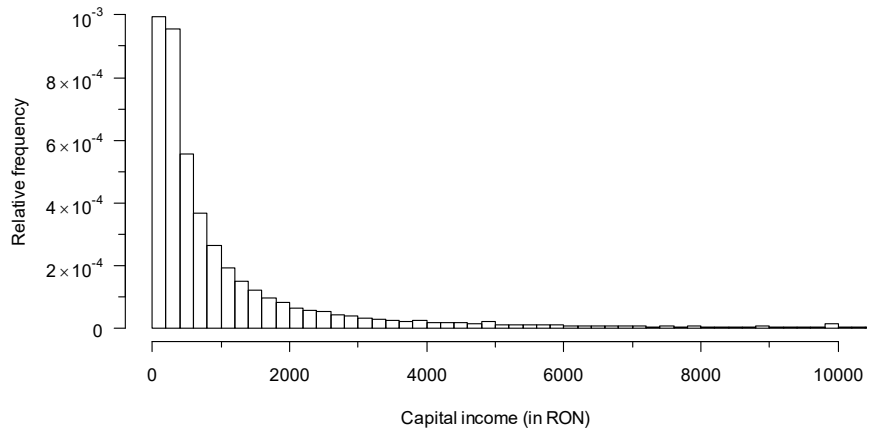

**(B)**

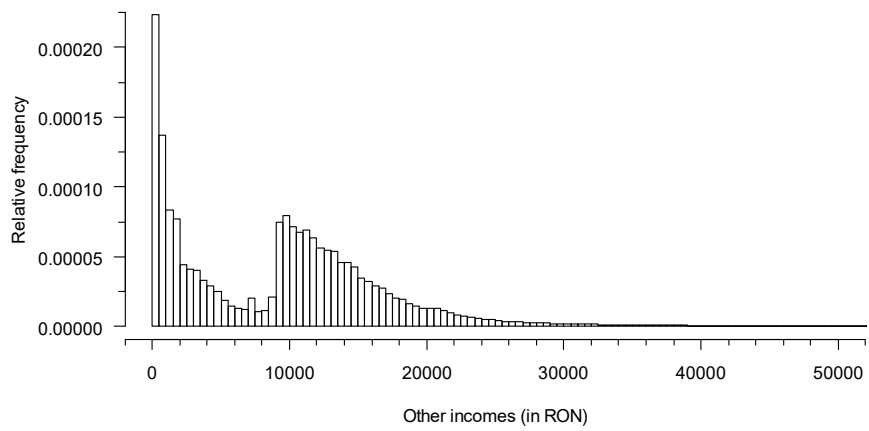

**(C)**

**Figure 4.** Histogram for wages, capital, and other incomes, across the population in 2014. (**A**) Wages; (**B**) Capital income; (**C**) Other incomes.

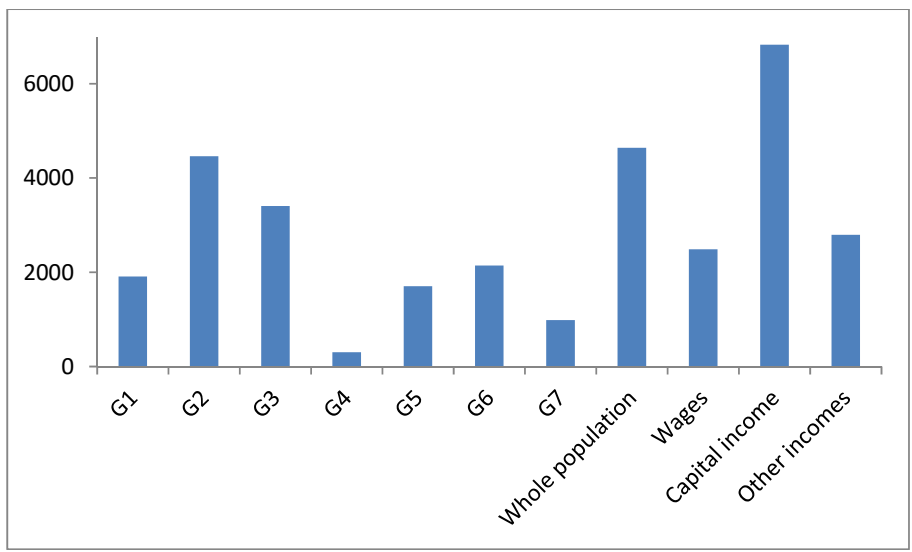

**Figure 5.** The ratio between top 1% incomes and bottom 99% incomes.

**Table 1.** Some descriptive statistics of the incomes series by sources of income.

| | Whole Population (Wages) | Whole Population (Capital Income) | Whole Population (Other Incomes) | Whole Population (Total Income) |
|---|---|---|---|---|
| Average | 20,623.12 | 13,752.05 | 12,017.18 | 20,826.43 |
| Standard deviation | 33,360.71 | 1,730,903 | 49,355.52 | 942,825.90 |
| Median | 12,400 | 598 | 9960 | 11,348 |
| Coefficient of variance (%) | 161.76 | 12,586.51 | 410.71 | 4527.07 |
| Top 10%/bottom 90% | 78.91 | 1005.03 | 208.40 | 387.27 |
| Top 1%/bottom 99% | 2487.23 | 6827.21 | 2795.66 | 4640.43 |
| Mean/Median | 1.66 | 23.00 | 1.21 | 1.84 |
| | **G1 (wages)** | **G2 (capital income)** | **G3 (other incomes)** | |
| Average | 18,574.47 | 9562.41 | 13,971.92 | |
| Standard deviation | 26,229.85 | 1149,281 | 60,275.29 | For G1, G2 and G3 the results for "Total income" are identical with those on different income source |
| Median | 12,035 | 470 | 10,428 | |
| Coefficient of variance | 1.41 | 12,018.75 | 431.40 | |
| Top 10%/bottom 90% | 65.94 | 713.38 | 183.14 | |
| Top 1%/bottom 99% | 1912.47 | 4460.42 | 3403.67 | |
| Mean/Median | 1.54 | 20.35 | 1.34 | |
| | **G4 (wages and other incomes)** | | | |
| Average | 14,759.89 | | 5009.29 | 19,769.19 |
| Standard deviation | 20,776.91 | | 13,612.08 | 26,077.56 |
| Median | 9843 | | 1611 | 13,203 |
| Coefficient of variance | 140.77 | Not the case | 271.74 | 131.91 |
| Top 10%/bottom 90% | 88.06 | | 169.03 | 27.48 |
| Top 1%/bottom 99% | 2182.35 | | 1358.48 | 305.33 |
| Mean/Median | 1.50 | | 3.11 | 1.50 |
| | **G5 (wages and capital income)** | | | |
| Average | 37,104.79 | 22,572.58 | | 59,677.37 |
| Standard deviation | 59,334.21 | 468,643.80 | | 473,716.70 |
| Median | 21,979 | 703 | | 27,973 |
| Coefficient of variance | 159.91 | 2076.16 | Not the case | 793.80 |
| Top 10%/bottom 90% | 54.45 | 1635.26 | | 69.82 |
| Top 1%/bottom 99% | 1564.80 | 10,661.56 | | 1705.22 |
| Mean/Median | 1.69 | 32.11 | | 2.14 |

| G6 (capital and other sources of income) | | | |
|---|---|---|---|
| Average | | 9198.97 | 13,612.48 | 22,811.45 |
| Standard deviation | | 1,816,826 | 38,784.26 | 1,817,392 |
| Median | | 666 | 12,501 | 14,049 |
| Coefficient of variance | Not the case | 19,750.32 | 284.92 | 7967.02 |
| Top 10%/bottom 90% | | 629.44 | 78.26 | 98.62 |
| Top 1%/bottom 99% | | 4897.98 | 1046.05 | 2142.05 |
| Mean/Median | | 13.81 | 1.09 | 1.62 |
| G7 (wages, capital and other incomes) | | | |
| Average | 26,545.61 | 24,924.32 | 11,244.93 | 62,714.87 |
| Standard deviation | 49,203.57 | 4,204,152 | 40,777.57 | 4,204,975 |
| Median | 14,400 | 825 | 4270.5 | 28,573.50 |
| Coefficient of variance | 185.35 | 16,867.67 | 362.63 | 6704.91 |
| Top 10%/bottom 90% | 104.36 | 1774.50 | 1328.07 | 54.01 |
| Top 1%/bottom 99% | 2975.03 | 13,926.05 | 1509.16 | 982.31 |
| Mean/Median | 1.84 | 30.21 | 2.63 | 2.19 |

**4. Breakdown by Disjoint Groups**

In order to measure the inequality of income distribution on each group, we calculate the Theil index. We consider the incomes earned by persons in a group $G_i$ is $v_{ij}$ and this group has $n_i$ persons. Under this condition, the income series for this group is represented by the vector $v_i = \left(v_{i1}, \dots, v_{in_i}\right), i = 1, \dots, 7$. The total income for each group is $VT_i = \sum_{j=1}^{n_i} v_{ij}$, the mean income of the group being denoted by $\mu_i = VT_i / n_i$, and the Theil index can be formulated as [37],

$$T_i(v) = \frac{1}{n_i} \sum_{j=1}^{n_i} \frac{v_{ij}}{\mu_i} \log \frac{v_{ij}}{\mu_i} \qquad (2)$$

The results obtained for the seven groups and the entire population are presented in Table 2.

**Table 2.** Theil indices computed for incomes at the level of entire population and for seven groups.

| Population | G1 | G2 | G3 | G4 | G5 | G6 | G7 | Whole Population |
|---|---|---|---|---|---|---|---|---|
| Theil Index | 0.47 | 3.73 | 0.88 | 0.42 | 1.06 | 1.56 | 1.95 | 1.18 |

In the following, we will analyze how the degree of the inequality of income distribution of the entire population is explained by two factors: the inequality within each group of persons and the differences between the seven groups of population. The seven groups being disjoint, the inequality of incomes explained by the Theil index can be decomposed as follows [37]:

$$T(v) = \frac{1}{n} \sum_{i=1}^{7} \sum_{j=1}^{n_i} \frac{v_{ij}}{\mu} \log \frac{v_{ij}}{\mu} = \sum_{i=1}^{7} \frac{VT_i}{VT} T_i + \sum_{i=1}^{7} \frac{n_i}{n} \left[\frac{\mu_i}{\mu} \log \frac{\mu_i}{\mu}\right] = T_{WI} + T_{BI} \qquad (3)$$

In (3), the Theil index computed for the entire population can be decomposed in two components:

(i) The first term ($T_{WI}$) measures the inequality of income distribution as a result of the differences concerning the distribution of income in the seven groups. For each group inequality of income distribution is calculated by Theil indices and for all groups we evaluate this part of $T(v)$ to multiply the Theil indices calculated at the group level by weighted arithmetic mean of all income;

(ii) The second decomposition term in relation (3), which is denoted by $T_{BI}$, quantifies the part of the inequality of distribution of population incomes due to the differences of income distribution that exist between the seven groups. This term is a Theil index calculated for the average income at the level of the groups and using as a relative frequency the structure of the population on the seven groups from which the population is constituted.

Table 3 shows the results of the decomposition of the inequality of income distribution based on the inequality of income distribution on each of the seven groups of people using (3).

**Table 3.** Decomposing the inequality of total income distribution across groups of people.

| | Theil Index | Share (%) of the Inequality Measured by the Theil Index of Each Group |
|---|---|---|
| G1 | 0.47 | 13.91 |
| G2 | 3.73 | 16.89 |
| G3 | 0.88 | 11.52 |
| G4 | 0.42 | 2.71 |
| G5 | 1.06 | 17.66 |
| G6 | 1.56 | 12.44 |
| G7 | 1.95 | 12.21 |
| Whole population | 1.18 | 100 |
| Within groups | 1.03 | 87.34 |
| Between groups | 0.15 | 12.66 |

**5. Decomposing the Inequality of Income Distribution by Income Sources**

For the G4, G5 and G6 groups each person can have two sources of income and for G7 three sources of income. For instance, a person belonging to G5 group has as income sources wages and capital income. Under these conditions, the inequality of income distribution within this group is determined by the distribution of the income of the persons in the group on each of the two sources of income, as well as by the distribution of the total income of the group on the two categories of income.

For the groups with two income sources, the total income ($VT$) is considered to be the sum of two income categories: $VT_j = X_j + Y_j, j = 1, \ldots, n$. It is denoted by $T_X, T_Y$ Theil indices calculated for the distributions of the two variables $X$ and $Y$, respectively. For the decomposition of the inequality of the distribution of the total incomes, relation (3) cannot not applied. In this case, the Theil index calculated for total income is the sum of three terms:

$$
\begin{aligned}
T_{VT} = \frac{1}{n}\sum_{j=1}^{n} \frac{VT_j}{\mu} \log \frac{VT_j}{\mu} \\
= \frac{\mu_x}{\mu} T_x + \frac{\mu_y}{\mu} T_y) + (\frac{\mu_x}{\mu} \log \frac{\mu_x}{\mu} + \frac{\mu_y}{\mu} \log \frac{\mu_y}{\mu}) \\
+ (-(\sum_{j=1}^{n}(\frac{x_j}{VT} \log \frac{x_j}{VT_j} + \frac{y_j}{VT} \log \frac{y_j}{VT_j}))
\end{aligned}
\tag{4}
$$

The first term in this relationship is a weighted arithmetic mean of the Theil indices calculated for the two variables. The weights are the ratios of income on each source of income and the total income. This term measures the income inequality due to the inequality of income distribution for each income source. In this case, we calculated the Theil index for each of the data series.

The second term measures income inequality due to distribution of income by the income sources. This term is a Theil index calculated on the basis of the distribution of income on the two sources of income.

The third term represents the correlation between the two categories of income that influence the inequality of income distribution among a population.

We present below the breakdown of the Theil index calculated for the total income of individuals of a population if they are formed on the basis of multiple sources of income. We denote by $V_j, j = 1, \ldots, n$ the income of the person formed on the basis of $m$ income sources. In this case, $V_j = X_{1j} + \cdots + X_{mj}, j = 1, \ldots, n$. The Theil Index of the entire population breaks down as follows:

$$
\begin{aligned}
T_{VT} = (\frac{\mu_1}{\mu} T_1 + \cdots + \frac{\mu_m}{\mu} T_m) + (\frac{\mu_1}{\mu} \log \frac{\mu_1}{\mu} + \cdots + \frac{\mu_m}{\mu} \log \frac{\mu_m}{\mu}) \\
+ (-\sum_{i=1}^{m}\sum_{j=1}^{n_i} \frac{x_{ij}}{VT} \lg \frac{x_{ij}}{VT_j}) = T_{WG} + T_{BG} + \Delta_{COR}
\end{aligned}
\tag{5}
$$

where $\mu_i$—represents the average income per person for the income symbolized by $X_i, i = 1, \ldots, m$; $\mu$—average income per person irrespective of the source of income; $T_i, i = 1, \ldots m$—Theil index for the income series $X_i$; $VT_j$—total income of a person; $x_{ij}$—the income earned by a person having $X_i$. as a source of income.

In relations (4) and (5), three terms are identified:

- The first relationship measures the inequality of distribution of total incomes of the population due to the differences that exist in the distribution of income distributed by each income source. In this case, the Theil indices computed on the series of data constituted by income sources are multiplied by the weights of the total income from each income category in the total income of the population ($T_{WG}$);
- The second term quantifies the differences that exist between people's income categories. This term is computed as the difference between the maximum entropy and the entropy of distributing the total income of the population by income sources ($T_{BG}$);
- The latter term is a rest that quantifies the effect of interaction between income distribution on each income category and total income distribution across the population ($\Delta_{COR}$).

In the case of the decomposition of the Theil index used in the situation where the population was composed of disjoint groups, there is no longer the interaction factor that is included in the relations (4) and (5).

Table 4 presents the results obtained by applying decomposition (5) to the groups in which individuals earn the total income from two or three sources of income:

**Table 4.** The decomposition of Theil index using (5).

|  | Theil Index | Contribution to Total Inequality (%) |
|---|---|---|
| G4: Wages | 0.53 | - |
| Other incomes | 0.97 | - |
| Whole population | 0.42 | 100.0 |
| $T_{WG}$ | 0.64 | 154.3 |
| $T_{BG}$ | 0.13 | 30.6 |
| $\Delta_{COR}$ | −0.35 | −84.9 |
| G5: Wages | 0.56 | - |
| Capital income | 3.12 | - |
| Whole population | 1.06 | 100.0 |
| $T_{WG}$ | 1.53 | 144.0 |
| $T_{BG}$ | 0.03 | 2.8 |
| $\Delta_{COR}$ | −0.50 | −46.8 |
| G6: Other incomes | 0.34 | - |
| Capital income | 4.43 | - |
| Whole population | 1.56 | 100.0 |
| $T_{WG}$ | 1.99 | 127.3 |
| $T_{BG}$ | 0.02 | 1.2 |
| $\Delta_{COR}$ | −0.45 | −25.8 |
| G7: Wages | 0.66 | - |
| Capital income | 5.29 | - |
| Other incomes | 1.00 | - |
| Whole population | 1.95 | 100.0 |
| $T_{WG}$ | 2.56 | 131.0 |
| $T_{BG}$ | 0.06 | 3.1 |
| $\Delta_{COR}$ | −0.67 | −34.1 |

## 6. Conclusions and Discussion

We have computed the income inequality in Romania taking account of the different modes of income (wages, capital, etc.) for more than 11 million individuals for the year 2014.

The data has been analysed using entropy methods and characterised using the Theil index.

This leads to the identification of three components. The first computed on the series of data constituted by income sources is multiplied by the weights of the total income from each income category in the total income of the population. The second quantifies the differences that exist between the population's income categories and is computed as the difference between the maximum entropy and the entropy of distributing the total income of the population by income sources. The third one is a remainder that quantifies the effect of interaction between income distribution on each income category and total income distribution across the population.

We believe entropy methods of the kind proposed here offer a new and interesting way forward for the analysis of income distributions. In a future paper, we will re-examine the data using the Tsallis entropy now widely used to characterise economic phenomena (See [48,49]).

**Acknowledgments:** The authors would like to thank Roxana Herteliu-Iftode, who checked the English for the later version of our manuscript.

**Author Contributions:** Bogdan Oancea prepared the data series, wrote the R code and performed the statistical data analysis. Tudorel Andrei, Bogdan Oancea, Peter Richmond, Gurjeet Dhesi and Claudiu Herteliu wrote the paper.

**Conflicts of Interest:** The authors declare no conflict of interest.